# Modeling how and why aquatic vegetation removal can free rural households from poverty-disease traps

Molly J Doruska[a], Christopher B Barrett[a,b], Jason R Rohr[c]

January 2024

[a] Charles H. Dyson School of Applied Economics and Management, Cornell University, Ithaca, NY, USA
[b] Jeb E. Brooks School of Public Policy, Cornell University, Ithaca, NY, USA
[c] Department of Biological Sciences, Environmental Change Initiative, Eck Institute of Global Health, University of Notre Dame, Notre Dame, IN, USA

*Corresponding author: mjd438@cornell.edu

**Abstract**

*Background*

Infectious disease can reduce labor productivity and incomes, trapping subpopulations in a vicious cycle of ill health and poverty. Efforts to boost African farmers' agricultural production through fertilizer use can inadvertently promote the growth of aquatic vegetation that hosts disease vectors. Recent trials established that removing aquatic vegetation habitat for snail intermediate hosts reduces schistosomiasis infection rates in children, while converting the harvested vegetation into compost boosts agricultural productivity and incomes. Our model illustrates how this ecological intervention changes the feedback between the human and natural systems, potentially freeing rural households from poverty-disease traps.

*Methods*

We develop a bioeconomic model that interacts an analytical microeconomic model of agricultural households' behavior, health status and incomes over time with a dynamic model of schistosomiasis disease ecology. We calibrate the model with field data from northern Senegal.

*Findings*

We show analytically and via simulation that local conversion of invasive aquatic vegetation to compost changes the feedbacks among interlinked disease, aquatic and agricultural systems, reducing schistosomiasis infection and increasing incomes relative to the current status quo, in which villagers rarely remove vegetation.

*Interpretation*

Aquatic vegetation removal disrupts the poverty-disease trap by reducing habitat for snails that vector the infectious helminth and by promoting production of compost that returns to agricultural soils nutrients that currently leach into surface water from on-farm fertilizer applications. The result is healthier people, more productive labor, cleaner water, more productive agriculture, and higher incomes.

*Funding*

This research was supported by a National Science Foundation grants DEB-2109293 and BCS-2307944.

**Research in Context**

*Evidence before the study*

We searched Google Scholar, EconLit, and other reference databases for research articles on poverty-disease traps published in English between March 1, 2021 and October 19, 2023. Our search terms included "poverty trap", "poverty disease trap", "poverty and infectious disease", "economics of infectious disease", and combinations of these terms. The current literature on poverty-disease traps mainly focuses on the aggregate (macro) scale of communities and countries, assuming a reduced form link between income and infection. A parallel, but thus far unlinked, literature on household-level poverty traps starts from microeconomic foundations of individual choices and outcomes. That literature largely ignores the feedback between socioeconomic behaviors and natural systems, focusing instead on how poverty traps arise from financial market failures and/or poor human nutrition. A separate literature exists on the disease ecology of schistosomiasis that abstracts away from human behaviors that lead to heavy disease exposure in some, but not all, communities.

*Added value of this study*

We develop a model that takes the poverty-disease traps literature to the micro scale, connecting a pre-existing disease ecology model of schistosomiasis infection dynamics to an analytical microeconomic model of an agricultural household optimally choosing its behaviors subject to environmental and market constraints. By rooting the poverty-disease trap in a structural microeconomic model of household decision-making, and by introducing a proper model of natural dynamics into an economic model of poverty traps, we integrate parallel literatures, providing a foundation for more precise exploration of the structural underpinnings of poverty traps based on human-nature interactions in general and of poverty-disease traps more specifically. This analytical model also provides a theory-based, but numerical and structural explanation for why a novel ecological intervention to clear aquatic vegetation from water points succeeds in dramatically reducing schistosomiasis infection rates while boosting agricultural productivity.

*Implications of all the available evidence*

Understanding the structural and interconnected behaviors of the human and natural systems that describe rural areas opens up powerful opportunities to design and test interventions that might free communities from poverty-disease traps, as this specific intervention in northern Senegal does, and to understand the conditions under which such interventions might succeed.

**Introduction**

Rural populations in low and middle income countries suffer relatively high infectious disease prevalence and low agricultural productivity, which jointly result in low incomes that can reinforce those conditions, resulting in a poverty-disease trap.[1-10] Efforts to intensify agricultural production and break out of the trap too often fail when inadequate attention is paid to how human behaviors interact with the dynamics of the natural ecosystems that support rural peoples' livelihoods, for example, when increased fertilizer use inadvertently aggravates infectious disease exposure.[11,12] Sustainably improving the livelihoods of millions of poor rural people requires structural understanding of the potential feedbacks among agricultural production, disease ecology, and rural households' behaviors and well-being.

One example of a poverty-disease trap involves schistosomiasis, a neglected tropical disease that currently infects more than 200 million people around the globe, with 800 million people at risk of infection.[13-15] Schistosomiasis is caused by a snail-hosted flatworm. Snails infected with schistosomes inhabit aquatic plants in freshwater habitats (lakes and rivers). These snails release larval schistosomes into the water, which then penetrate the skin while people perform daily activities, like bathing, washing clothes or swimming.[16,17] Adult worms settle in the veins surrounding the gastrointestinal (*Schistosoma mansoni*) or urinary (*Schistosoma haematobium*) tract of infected individuals. The eggs released by the worms trigger chronic inflammatory responses causing several ailments including, but not limited to, loss of tissue function, resulting in reduced physical energy – and thus labor supply– among adults and stunted growth and learning deficits among children.[18-20] Conventional methods to control schistosomiasis rely on mass deworming, whereby all children and/or adults within a village receive deworming medication to clear current infections. Mass deworming does not clear snails and schistosomes from the water sources, thus reinfection occurs quickly.[21,22] While mass deworming can generate large, transitory reductions in human infection levels, reducing long-term cycles of schistosomiasis infection and reinfection requires strategies that target different parts of the infection cycle.[22-24]

Recent field trials revealed that schistosomiasis in schoolchildren can be reduced by removing aquatic vegetation that serves as the habitat for snail intermediate hosts, complementing infection control through deworming.[12] Researchers converted this aquatic vegetation into compost and livestock feed, which improved agricultural production and lowered agricultural input costs. Aquatic vegetation removal for infectious disease control or the production of agricultural inputs is not currently widely practiced in the northern Senegal study region or elsewhere. It is therefore important to understand why this practice works and whether it might offer a transferable method for escaping from poverty-disease traps by offering households an economic incentive to remove aquatic vegetation, thereby reducing schistosomiasis exposure while simultaneously boosting agricultural productivity and household incomes.

We develop a bioeconomic model to examine the relationship among agricultural production, poverty, and disease in northern Senegal and to explore if and why aquatic vegetation removal can break poverty-disease traps. We start with a classic non-separable microeconomic model of agricultural household behavior[25] and connect it to a disease ecology model of schistosomiasis dynamics,[26] linking the models through household decisions about labor allocation, aquatic vegetation harvest, and fertilizer application, decisions that affect both agricultural outcomes and the underlying aquatic ecosystem and thereby (indirectly) the probability of human infection. Existing macro-scale models of poverty-disease traps necessarily abstract away from individual-level incentives and behaviors,[5,8,9] relying on reduced form associations at the population scale. We instead follow a tradition of structural microeconomic models that explicitly link human behaviors to the dynamics of natural phenomena.[27-30] A structural microeconomic model enables us to identify the conditions under which households might voluntarily undertake aquatic vegetation removal, those under which vegetation removal may suffice to control schistosomiasis transmission, and how such incentives and outcomes vary with household attributes, such as farm size.

Our results highlight two key feedback loops households face. First, under the status quo, with no aquatic vegetation removal, we see explicitly how a poverty-disease trap emerges. Vegetation growth remains unchecked by households, boosting schistosomiasis infection rates that reduce household labor supply, which in turn reduces the time allocated to agricultural production and thus overall incomes. If, however, households implement a very simple intervention, clearing the water access point of invasive weeds that host the snails that vector the schistosomes, infections plummet and labor supply, agricultural productivity and incomes increase, helping to break the poverty-disease trap.

Second, fertilizer run-off provides key nutrients that foster aquatic plant growth, reducing the effectiveness of aquatic vegetation removal and thereby allowing snails and infection to persist, making it more challenging for households to break the poverty-disease trap. This reveals an under-recognized tradeoff in agricultural development efforts; while fertilizer use increases agricultural output, it can also indirectly promote infectious disease exposure, with analytically ambiguous effects on health, incomes and living standards, much like pesticides.[31] Together, these main results demonstrate the importance of understanding and considering structural feedbacks when proposing interventions to improve livelihoods and enable escapes from poverty-disease traps.

**Bioeconomic Model**

The bioeconomic model has two submodels. The first describes the disease ecology dynamics, more specifically, how the schistosome, aquatic vegetation, and snail populations interact, and relates these populations to human infections. The second, an agricultural household submodel, describes how utility maximizing households make decisions about how to allocate their land, labor, and income.

The household's problem is a variant of the non-separable agricultural household model in which consumption and production decisions become inextricably linked by multiple market failures that typically characterize poor rural villages like those in our setting.[25] The economic model begins with a representative household that maximizes utility, defined over consumption of food, an aggregate non-food household good, leisure, and the health status of household members. We assume that utility is well-defined, increasing and concave in all its arguments. We model the household's nutrient intake via food consumption. The health production function is Cobb-Douglas for food consumption and the fraction of household members infected downscales the health status variable as the fraction infected increases. Health status increases with food consumption, representing the value of more nutrient intake. The household can only influence health status through more food consumption or a lower infection prevalence; one cannot buy good health. Because aquatic vegetation is a common pool resource, there is no market for aquatic vegetation, either in the water or as harvested vegetation. The multiple market failures in health status and aquatic vegetation together create non-separability between the household's production and consumption decisions. To simplify the model, we also assume no market exists for land rentals or sales and from cash labor markets as land or labor transactions are uncommon in the study area. Households allocate their time among cultivating food, harvesting aquatic vegetation, and leisure and commit their land to their own agricultural production. These assumptions do not qualitatively change model outcomes.

If households choose to harvest aquatic vegetation, they turn it into compost, which increases agricultural productivity.[12] Households produce food using land, labor, fertilizer, and compost from harvested aquatic vegetation. Recent experimental evidence finds that compost and urea fertilizer are virtually perfect substitutes.[12] Harvesting vegetation only requires labor.[1] The household employs a constant elasticity of substitution (CES) food production function while aquatic vegetation harvest follows Cobb-Douglas production technology.[2] More details on the agricultural household model are in the supplementary materials.

To simulate the status quo ex ante, we also present a simplified version of the model without aquatic vegetation harvest, in which households cannot use labor to harvest aquatic vegetation to produce compost. Our core comparisons thus simulate the equilibrium effects of making villagers aware of the prospective value of composting harvested aquatic vegetation.

---

[1] While it requires a pit to convert vegetation into compost, we assume there exists sufficient unused, free land within the village such that land availability does not constrain compost production.
[2] Labor is the only input to harvest vegetation, so there is no need for a CES to allow for substitution among inputs.

The disease ecology model tracks the populations of aquatic vegetation (*Ceratophyllum*, $N$), miracidia (larval schistosomes that infect snails, $M$), infected and susceptible snails ($I_2$ and $S_2$), cercariae (larval schistosomes that infect humans, $P$), and infected and susceptible humans ($I_1$ and $S_1$). We adapt an existing schistosomiasis disease ecology model[26] to fit the Senegalese context and down-scale the parameters from a large community to one that matches the household-level simulations. Additional details on the disease ecology model are in the supplemental materials.

Relative to the human lifespan, the schistosomiasis infection cycle is relatively short. Cercariae live around 10 hours, miracidia live around 25 hours, and snail infections last around 100 days.[32] Very few or none of the existing cercariae or miracidia population will survive over the course of the year, which creates a challenge to match timescales across the household and disease ecology submodels. One could convert the continuous time disease ecology submodel to discrete time to match the household submodel through significant linearization and assumptions about annual changes in miracidia, cercariae, and snail populations. But that can cause meaningful aggregation errors. We therefore instead use a continuous time disease ecology submodel that better matches the timeline of the schistosomiasis infection cycle. We simulate annual changes by simulating the system of differential equations forward 365 days, where all parameters are given in daily rates. We then export the annual output to the discrete time household model that operates at annual time steps.

The disease ecology submodel and the household submodel link to one another in two ways. The first is through the infection status of the household, which directly affects household utility and impacts the household's labor availability. The second is through the household's use of urea fertilizer and its aquatic vegetation harvest, each of which changes the vegetation population within the water source.

The disease ecology submodel provides population estimates of infection, which we scale down to individual- and household-level infection rates through stochastic infection realizations drawn from an independent Bernoulli distribution for each household member at the start of each time period. The distribution's mean is the infection rate predicted by the disease ecology submodel, the population infection prevalence. After the first period, we also take random draws for curing infection, where the mean of the Bernoulli random variable was set at 0.25, which captures the fact that households in this region experience sporadic mass deworming campaigns.[33]

Since each individual household is only one small part of a village and villages only access a small portion of the entire aquatic system, these households do not individually influence the disease ecology submodel. Since household behavior does not individually impact disease ecology, the household does not consider the equations of the disease ecology submodel in its own optimization. In this way, the household solves a series of static, single period optimization problems as in prior bioeconomic models.[27,[3]] In this framework, the disease ecology submodel

---

[3] Any of several justifications exist to follow this approach. Households cannot fully control the decisions of all household members, such as parents telling their children to stay out of the water but children not listening, thus the natural dynamics escape household control. Or households might not fully understand the evolution of the disease ecology submodel as given in the equations that connect vegetation, miracidia, cercariae, snails, and humans. Each of these is likely true to some degree, allowing us to avoid the unrealistic and computationally task of modeling a household that monitors all seven populations in the disease ecology submodel as state variables. That would require

shows how the state and the average infection rate change over time. In each period, we solve the household's static optimization problem and then use the household's choices to determine the stock of aquatic vegetation and the realizations of infection to determine the current infection prevalence. With these new starting populations, we simulate the disease ecology model one year forward to give the state of infection in the next time period. The model is then solved by the following iterative process for each period in the simulation:

1. We use Bernoulli random draws to realize household infection;
2. The household solves their static problem by allocating its time and money to maximize its period-specific utility;
3. Using the realizations of infection and the household's decisions, we calculate the current aquatic vegetation population and the current number of infected and susceptible individuals. We use these starting values and simulate the disease ecology submodel forward one year and calculate the vegetation population and village infection rate in the following period;
4. Repeat from step one for 20 annual periods.

Additional details on the linkages between submodels are in the supplemental materials.

We limit simulations to 20 years to explore the within-generation results of the model to see what happens when aquatic vegetation harvest is introduced, in particular, if vegetation harvest becomes a sustained behavior, resulting in new levels of (reduced) equilibrium infections and (higher) household incomes. This time frame is long enough to capture any short-term changes in the equilibrium level of schistosomiasis infection but allows us to abstract away from long-

---

significant discretization or a large reduction in the number of states to solve given the curse of dimensionality in optimal control problems.

term changes, including through impacts on children's educational attainment, or in human fertility behaviors that would further complicate the model.

We simulate the model in Julia 1.6.2 and aggregate and analyze the model output in Stata 16. For each household type, we conduct 1,000 stochastic simulations to capture different optimal paths based on the realized random infection draws. Household types are determined by land holdings, which are set at the 25$^{th}$, 50$^{th}$, and 75$^{th}$ percentiles of land holdings in the Saint Louis and Louga regions based on the Harmonized Survey on Household Living Standards in Senegal 2018-2019 (Table S1).[34] Land holdings are proxies for wealth in this context and these simulations. Comparisons across land holding types give insight into how wealth levels impact the optimal decisions of the household. We track the following key outcome variables: household labor availability, labor allocated to food production, leisure, fertilizer use, the vegetation load in the water source, the household's level of infection, and the household's income. We then take the median of 1,000 simulations for each outcome at each time period for each household land endowment.

To begin, we eliminate the household's option to remove vegetation and produce compost by mechanically setting the marginal product of labor in aquatic vegetation harvest to zero. This lets us model how households currently behave and establish starting levels of infection and income under current conditions.

**Results**

When households do not harvest the aquatic vegetation, it remains stable at carrying capacity (figure 1A). Household infection reaches a high steady state (figure 1B, C). Households spend most labor on their farm and use moderate amounts of fertilizer in food production. High infection rates limit labor supply, leading to low income and a poverty-disease trap. These patterns are very similar across the wealth distribution.

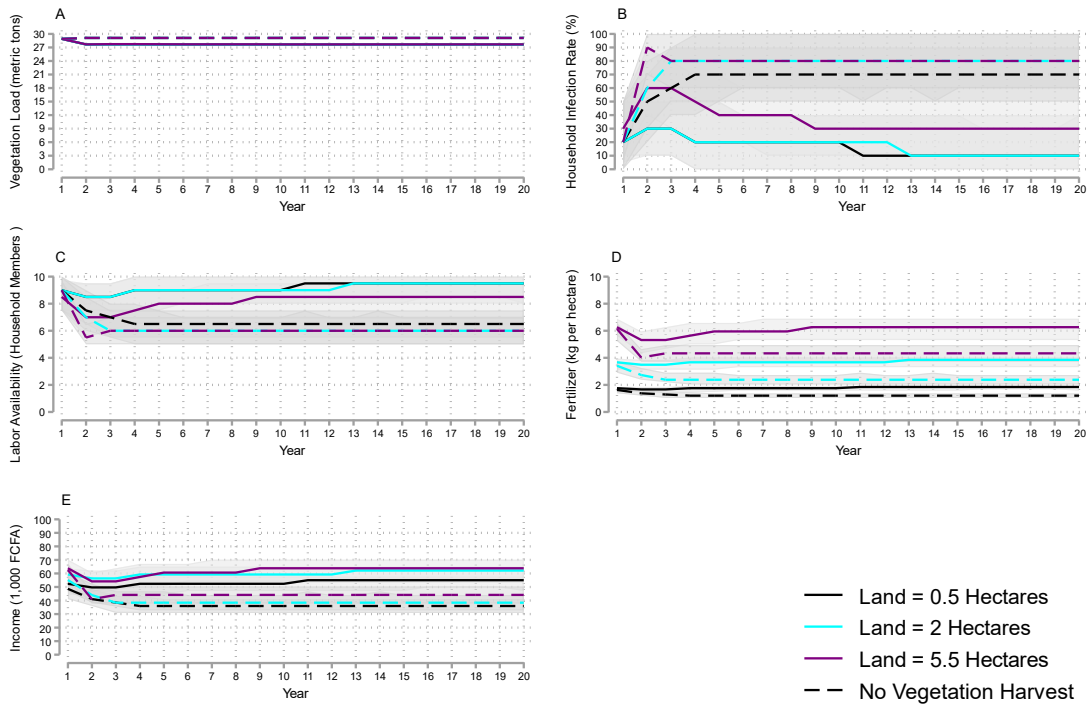

**Figure 1. Median vegetation load, infection rate, labor availability, fertilizer use, and income for simulations with and without vegetation harvest.** Panel A plots the median aquatic vegetation stock (population) in metric tons across 1,000 20-year simulations for three different household land endowments with (solid lines) and without (dashed lines) vegetation harvest. Aquatic vegetation load represents the size of the snail habitat within the village water access point used by the household. Shaded areas represent the 5⁻95 percent centered confidence band. Based on scale and precision, not all shaded areas are visible. Panel B shows the median household infection rate (the number of infected individuals divided by total number of household members). Panel C reports median labor availability from the 10-person household size maximum. Panel D displays median fertilizer use in kgs per hectare, and Panel E reports the median income in FCFA1,000. Medians and percentiles are within each land endowment each time period across the 1,000 simulations.

Once the household can harvest aquatic vegetation, all household types allocate only a small fraction of their labor to that task, but with considerable impact. Households' clear some vegetation from the water source, leading to a stable vegetation level below the carrying capacity, consistent with field experimental data finding that 10 or fewer individuals could clear a village's water access points in a day.[12] Even with continued household fertilizer use, modest effort allocated to aquatic vegetation harvest maintains a reduced aquatic vegetation stock, driving down the household infection rate, especially for poorer households with low or moderate land endowments (figure 1D). Most household labor remains allocated to food production, but lower infection rates mean greater labor availability. This greater labor, in addition to the nutrients returned to the soil from the compost, leads to higher median incomes than in the baseline case without vegetation harvest (figure 1E). These results highlight that the attractive economic returns to compost created from the harvested aquatic vegetation[12] can help disrupt disease ecology dynamics, both reducing infection rates and boosting incomes in a favorable reinforcing feedback loop. The model helps us understand the underlying mechanisms that explain how and why the intervention seems to work.

The lack of smooth results in labor availability and household infections (figure 1) can be attributed to the stochastic process that generates household infections within the model.

Perhaps surprisingly, fertilizer use is higher when we allow for vegetation harvest. Since compost and fertilizer are substitutes, one might expect fertilizer use to decrease as farms begin harvesting aquatic vegetation. But such substitution effects are often dominated by income effects, as the prior household modeling literature has long established. Aquatic vegetation harvest increases incomes by increasing household labor availability and food productivity. Those higher incomes stimulate greater household food demand and relax financial liquidity constraints to fertilizer purchase. Thus the income effect is stronger than the substitution effect and fertilizer use increases when we introduce compost.

Additionally, our simulations are consistent with the empirical association of fertilizer use with infectious disease exposure[11,12] and therefore the optimal level of fertilizer for a household may depend on the level of infection. To test this directly, we re-ran the simulations starting with households at different infection rates and keeping all other model parameters the same. We calculated the median optimal first-year fertilizer use and plotted it across the different starting infection conditions (figure 2). Optimal fertilizer is negatively associated with infection rate, as predicted. The decrease is economically significant as very high levels of infection are associated with almost a 50% decrease in optimal fertilizer use compared with low infection levels.

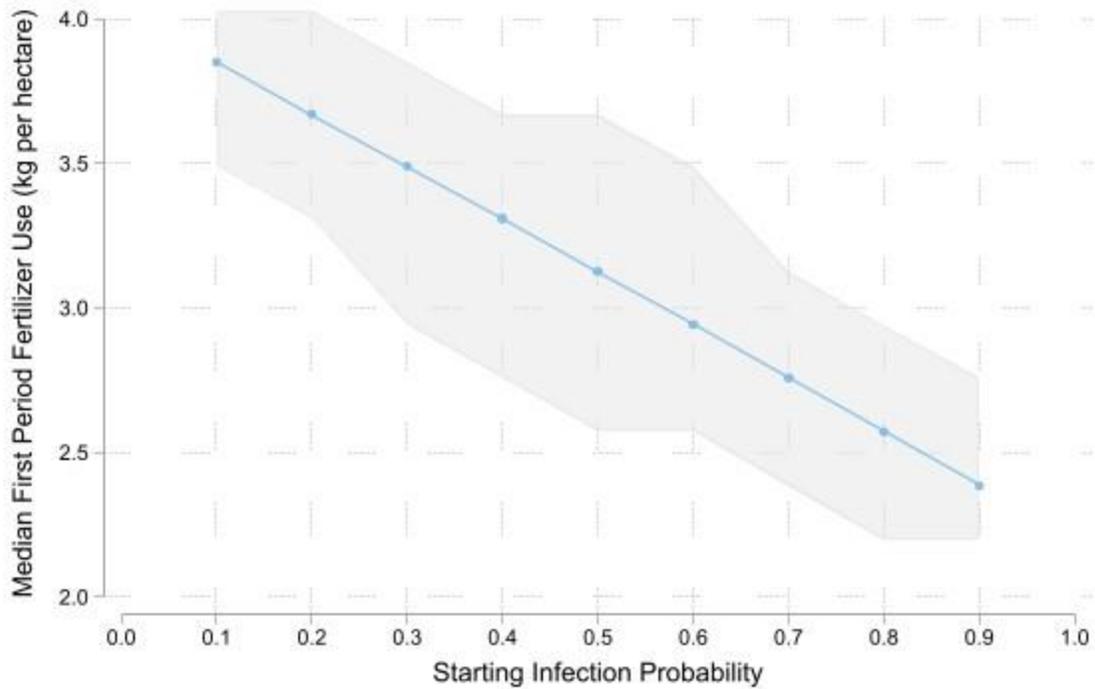

**Figure 2. Median first year optimal fertilizer use at different infection levels.** Figure 2 plots median fertilizer use in kgs per hectare in the first year across 1,000 20-year simulations for the median household land endowments with vegetation harvest. Shaded areas represent the 5-95 percent centered confidence band. Only the initial starting infection probability was modified. All other disease ecology submodel parameters remain the same.

**Sensitivity Analysis**

We explore the sensitivity of our results to the effect of fertilizer runoff on vegetation ($\rho$), the vegetation recolonization rate ($n0$), the vegetation growth rate ($r$), and the price of fertilizer ($p_u$). We also conduct a sensitivity analysis of the price of the household good ($p_h$).[4] For the sensitivity analysis, we focus on changes to parameters in the system and consider the median household land holding of two hectares.

---

[4] The ratio of prices governs the economic incentives households face, so changing the price of the household good implicitly changes the relative value of food.

The core model results described above are generally robust to changes in the effects of fertilizer runoff on vegetation growth, recolonization rate, and growth rate, and economic incentives modeled through changes in the price of fertilizer and the household good. Slightly higher levels of infection and lower labor availability result when the fertilizer runoff effect (figure 3) and vegetation recolonization rate increase (figure S1).

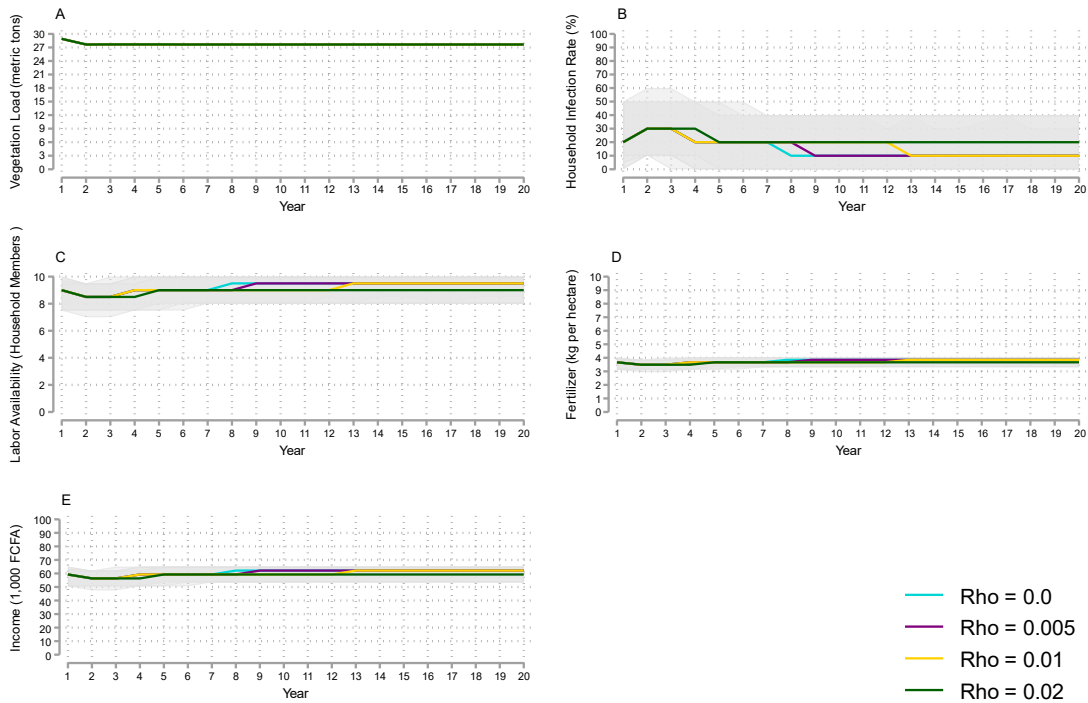

**Figure 3. Median vegetation load, infection rate, labor availability, fertilizer use, and income for the fertilizer effect sensitivity analysis.** Panel A plots the median aquatic vegetation stock (population) in metric tons across 1,000 20-year simulations for four different levels of feedback between fertilizer runoff and vegetation growth ($\rho$) and households with two hectares of land. Aquatic vegetation load represents the size of the snail habitat within the village water access point used by the household. Shaded areas represent the 5-95 percent centered confidence band. Based on scale and precision, not all shaded areas are visible. Panel B shows the median household infection rate (the number of infected individuals divided by total number of household members). Panel C reports median labor availability from the 10-person household size maximum. Panel D displays median fertilizer use in kgs per hectare, and Panel E reports the median income in FCFA1,000. Medians and percentiles are within each land endowment each time period across the 1,000 simulations.

At lower levels of vegetation growth, the vegetation stock is smaller, infection prevalence is lower, household labor availability is higher, and income is slightly improved (figure 4).

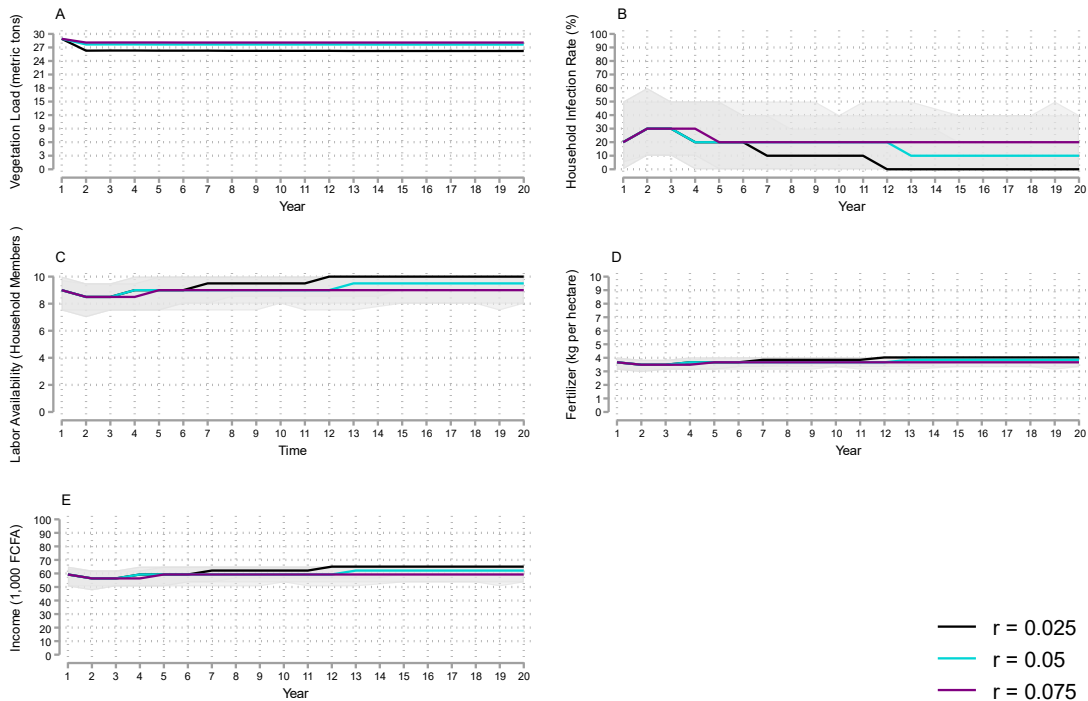

**Figure 4. Median vegetation load, infection rate, labor availability, fertilizer use, and income for the vegetation growth rate sensitivity analysis.** Panel A plots the median aquatic vegetation stock (population) in metric tons across 1,000 20-year simulations for three different levels of the vegetation growth rate ($r$) and households with two hectares of land. Aquatic vegetation load represents the size of the snail habitat within the village water access point used by the household. Shaded areas represent the 5-95 percent centered confidence band. Based on scale and precision, not all shaded areas are visible. Panel B shows the median household infection rate (the number of infected individuals divided by total number of household members). Panel C reports median labor availability from the 10-person household size maximum. Panel D displays median fertilizer use in kgs per hectare, and Panel E reports the median income in FCFA1,000. Medians and percentiles are within each land endowment each time period across the 1,000 simulations.

As expected, cheaper fertilizer leads households to use more of it, which results in modest increases in infection prevalence (figure S2). Finally, our results show no meaningful changes when the price of the household good changes (figure S3). Together, these results show that the patterns in our main results are consistent across a range of reasonable values for underlying agroecosystem and market conditions, and thus provide a robust structural way to capture the relationship between aquatic vegetation growth, the microeconomic decisions of households, and poverty and disease outcomes. The returns to compost in food production are routinely large enough to induce aquatic vegetation harvest if people are aware of the household benefits. However, the impact of aquatic vegetation harvest can be muted at higher levels of fertilizer use because vegetation growth spurred by fertilizer use offsets some of the gains made by harvesting aquatic vegetation.

**Discussion and Conclusion**

We developed a micro-structural model of a poverty-disease trap by linking a non-separable agricultural household model to one of schistosomiasis disease ecology dynamics through household labor availability, labor allocation choices, and optimal fertilizer use. The household-centered approach allows us to analyze how poverty-disease traps can exist under current conditions and how and why simple, low-cost interventions like aquatic vegetation harvest can help break those traps. Under the status quo, without aquatic vegetation harvest, infection prevalence is consistently high and household labor availability and income are steadily low. When we allow for vegetation harvest in the model, simulating what might happen after an agricultural extension and public health information campaign to promote aquatic vegetation removal, we see consistently lower infection levels and higher incomes. The effect of aquatic vegetation harvest is greater in combination with measures that reduce nutrient runoff that spurs aquatic vegetation regrowth. Continued household fertilizer use limits the gains for those with the highest land holdings, signaling that this seems a pro-poor intervention. Thus, aquatic vegetation harvest has the potential to allow households to reduce the cycle of schistosomiasis infection and reinfection that characterize the poverty-disease traps currently confronting many rural households in northern Senegal, and many other communities in the low-income tropics.

One limitation of our modeling strategy is that we only explore the representative household's choices, but water sources and water access points serve many households at one time. In the case of fertilizer use, one household's decision to use lots of fertilizer will inevitably impact the common water source, increasing the aquatic vegetation and schistosomiasis reservoir for all households who use that water source. This provides an opportunity for households to harvest more vegetation, but it also poses a greater infection risk due to other households' decisions. A natural extension of the current model would build out these interhousehold and spatial interactions into a dynamic general equilibrium bioeconomic model to trace out within-village spillover effects. Our results suggest that fertilizer use is an important piece of the system. Documenting these village externalities may prove helpful to fully understanding and tackling the poverty-disease trap.

While this model focuses on understanding how and why this intervention works in the specific context of the Saint Louis and Louga regions in northern Senegal, the principles of the intervention likely apply to other settings where schistosomiasis is endemic. *Ceratophyllum demersum*, the keystone aquatic vegetation species of interest in this model, is found throughout Africa and on every continent with endemic schistosomiasis.[17] Therefore, the aquatic vegetation removal model might apply to settings throughout the developing world, potentially benefitting millions who suffer from schistosomiasis infection. Furthermore, recent findings suggest that targeting snails, as through aquatic vegetation removal, is the most effective way to reduce schistosomiasis transmission.[26] We identify and structurally model a key potential mechanism to reduce infectious disease burdens in the low-income tropics, demonstrating the importance of understanding feedback loops between household economic decision making and the underlying natural environment, which has applications to other neglected tropical diseases and to other complex relationships between human and environmental systems.

Poverty-disease traps are widespread, thus understanding solutions is important. Research in Kenya finds significant impacts of deworming on child learning and that of their siblings[35,36] and labor market outcomes later in life after deworming.[37,38] Thus, there are likely large potential long-term benefits of aquatic vegetation removal not modeled nor discussed here given the twenty-year time horizon we impose in the modeling. Policy makers, community leaders and development agencies should consider aquatic vegetation removal as an effective form of schistosomiasis infection control that can also boost incomes and overall quality of life for millions of people.

**Data and code availability**

All code and simulated data can be found at https://github.com/mdoruska/Bioeconomic_Model.

**Acknowledgements**

For helpful comments and discussions, we thank Brian Dillon, Chris Haggerty, Nicolas Jounard, Kira Lancker, Shanjun Li, Sean Moore, Alex Perkins, Ivan Rudik, Alex Sacks, and seminar participants at Cornell University. Any errors are the authors' sole responsibility.

**Supplementary Materials**

**SI Appendix Text S1. More Information on the Literature and Context of the Study**

*Poverty-Disease Traps*

Poverty-disease traps, perhaps first mathematically modeled by Bonds and colleagues,[1] connect a classic susceptible-infected-susceptible (SIS) general disease model to income where key model parameters that define death, recovery, transmission, and general infection are functions of income, and income is a function of infection. This model, and expansions on it, [2-8] fall into two broad categories, those that maintain the basic feedback loop Bonds and colleagues employ[1] but add stochasticity or other refinements[7] and those that apply the idea of a poverty-disease trap to other modeling frameworks.[2,4-6,8]

The primary extension of the basic model in the literature comes from connecting a neoclassical macroeconomic growth model to a disease ecology model where capital accumulation depends on infection.[5,8] The systems exhibit evidence of poverty traps.[5,8] Such models lack a micro-economic foundation to explain individuals' decisions, however, which makes it somewhat difficult to understand the structural behavioral foundations of the reduced form relationships that underpin the model. Ngonghala and colleagues[6] develop 11 different versions of the basic neoclassical growth model that include up to three types of capital (human, physical, and biological) and populations of natural enemies, parasites, pests, and predators. Goenka and Liu[4] add public or private investment to control disease transmission to the macro-level neoclassical growth model. The authors find disease slows growth and makes poverty traps possible. These neoclassical growth models consider only larger aggregates of people: villages or countries.

A much smaller literature looks at individual or household decision making relating to malaria in Uganda[2] and Buruli ulcer.[3] Berthélemy and colleagues[2] use theoretical models to derive the infectiousness of malaria and then demonstrate under which conditions the spread of malaria might result in a poverty trap. Garchitorena and colleagues[3] model the individual or household with a Cobb-Douglas production function and they find that even with relatively low incidence of disease, as with Buruli ulcer, poverty-disease traps are possible, especially when areas start with high levels of poverty. However, these economic models do not include clear modelling of the tradeoffs faced by individuals making decisions. The authors instead model the decision to treat a disease with random draws based on exogenous probability distributions.

This paper uses the analytical base of the microeconomic behavior of a household that makes optimal decisions and faces trade-offs because of binding budget and time constraints. By explicitly depicting the primal structural problem that households face when making choices subject to constraints, we demonstrate not only *that* households can be trapped in poverty due to infectious disease exposure, but also can identify *why* and thus *how* one might change underlying behaviors and outcomes. Indeed, this framework allows us to consider the formal comparative statistics of the household's constrained optimization problem, to explore how changes in prices or quantities of goods impact the household's optimal decisions, in addition to simulate the system empirically to allow us to better identify feedback loops within the system and thus how and why a specific intervention works – or fails to work – to reduce infection and poverty rates.

*Senegalese Context*

The geographical context for this paper is the Senegal River Valley and the Saint Louis and Louga regions in northern Senegal. The 1988 construction of the Diama dam, near the mouth of the Senegal River, dramatically changed land use in the region, particularly along the shores of the Senegal River and Lac de Gueirs, the largest basin within the region.[9,10] The creation of irrigation canals and the subsequent desalination of the water expanded the habitat of *Bulinus* and *Biomphalaria* snails, the intermediate vector for schistosomiasis transmission. *S. mansoni* and *S. haematobium* are currently endemic within the region.[10] About 75% of school children within 16 villages in the region were infected with *S. haematobium*, a urogenital schistosomiasis infection, while 25% of school children were infected with *S. mansoni*, a colorectal infection. Many of the children infected with *S. mansoni* are also infected with *S. haematobium*.[11] Around 90% of cattle within the region were infected with *Schistosomiasis bovis* (a livestock variant of schistosomiasis), and many of the *S. haematobium* infections within humans in the region are *S. haematobium - S. bovis* hybrid infections.[10]

Villages within the region are small, typically with populations between 1,000 and 5,000 residents.[11] Households within this region are largely agricultural, predominately growing rice, millet, cowpea, and peanuts.[12] Other horticulture crops are commonly grown in smaller plots. Many households within these villages rely on surface water sources to wash clothes, bathe, and irrigate plots. There also is sugar cane production along the northern edge of Lac de Gueirs which contributes to significant fertilizer runoff and ecological concerns, particularly eutrophication, within the lake. Increased nutrient loading within the water source contribute to the growth of *Ceratophyllum demersum*, the aquatic vegetation that is the preferred habitat for snails, and thereby to increased schistosomiasis infection. [13, 14]

The 2018-2019 Harmonized Survey on Household Living Standards in Senegal collected by the West African Economic and Monetary Union (WAEMU) Commission[12],[5] reports that 30% of the household heads in the survey are female, and on average household size is large with over 10 members per household (Table S2). The average household head is 52 years old and over 85% of household heads are married. Literacy rates are low as just under a third of household heads can read and write in French. Just under 30% of households engage in rice cultivation, around 40% of households have irrigation on at least one of their plots, and 45% of households use fertilizer on at least one of their plots. Households devote just under 400 person days to working on their farm across all family members. Just over 40% of households hire outside labor to work on their farm and the average family hires outside labor for 23 person days. Conditional on households hiring any outside labor, households hire on average outside labor for almost 35 person days.

*Aquatic Vegetation Removal*

Transmission of schistosomiasis occurs through the intermediate vector of aquatic snails. The parasite enters the water source when an infected human or animal (especially, cattle) urinates or defecates in the water releasing schistosome eggs. Once in the water, the eggs release miracidia, the first parasitic larval stage that infects the aquatic snails. After four to six weeks in an infected snail, cercariae, a subsequent larval stage of the parasite, exit the snail. Humans become infected with *Schistosoma* spp. worms through water contact with cercariae that enter the body through the skin.[15]

---

[5] Source: WAEMU Commission, Harmonized Survey on Household Living Standards, Senegal 2018-2019. Ref. SEN_2018_EHCVM_v02_M. Dataset downloaded from www.microdata.worldbank.org on September 2, 2022.

The aquatic vegetation removal intervention modeled in this paper specifically looks to disrupt the infection cycle through reduced snail habitat. *Bulinus* and *Biomphalaria* snails live in the submergent vegetation, *Ceratophyllum demersum*, in the lakes and rivers of the region. This aquatic vegetation has a symbiotic relationship with the snail population and cercariae.[16] By removing the aquatic vegetation, snails lose their habitat and source of food reducing both the number of snails and the cercariae they release.

Previous experimental work in this region suggests that removing aquatic vegetation from freshwater sources can significantly reduce *S. mansoni* infection in children through decreased snail populations.[11] As such, this model focuses on *S. mansoni* gastrointestinal infection. Recent crop trials suggest that compost made from harvested vegetation increases onion and pepper yields offering a good substitute for fertilizer.[11] By producing compost from aquatic vegetation sourced from the system, nitrogen applied on the fields in the form of compost from this vegetation simply recycles nitrogen that already existed within the system. Vegetation removal thus has the possibility to close nitrogen loops within the region, both boosting agricultural productivity by reusing leached nutrients and reducing infection prevalence by reducing snail habitat.

**SI Appendix Text S2. Additional Details of the Bioeconomic Model**

*The Household's Problem*

Let $i$ denote each of the different goods a household consumes, produces, or uses as a production input. Let $q_i$ denote the quantity of goods produced or used as production inputs by the household. The household produces ($q_i \geq 0$) of food ($i = f$) using land ($i = d$), labor ($i = l_f$), fertilizer ($i = u$), and compost ($\omega q_v$). The household makes compost from harvested vegetation ($i = v$) and harvesting vegetation requires labor ($i = l_v$). Composting reduces the mass of harvested vegetation, so the fraction of harvested vegetation remaining as compost to use in food production is $\omega \in (0,1)$. Let $L_f = q_{l_f} + L_f^h$ be the total amount of labor used in the production of food and $L_v = q_{l_v} + L_v^h$ be the total amount of labor used to harvest vegetation. The household's production technology for food is then given by $F(L_f, q_d, q_u, \omega q_v)$ and the production technology for harvesting vegetation is $G(L_v)$.

Let $\mathbf{c}$ denote the vector of all consumption goods comprised of food ($i = f$), non-food household goods and services ($i = g$), and leisure ($i = l$). Let $H(I_1, S_1, c_f)$ denote the household's health status, which is an decreasing function of $I_1$, the number of infected individuals in the household, and $S_1$, the number of not infected (susceptible) individuals in the household,[6] and increasing in $c_f$, food consumption. We denote household utility as $U(\mathbf{c}, H)$.

Each household has endowments of labor $e_l$ and land $e_d$ in each time period. Each household member has one unit of labor; however, infection reduces the labor availability of an individual to $\tau$ where $0 \leq \tau < 1$. Infection reduces nutrient absorption from food and results in less labor productivity overall, effectively reducing the labor availability of infected individuals. The labor available to the household $a_l$ is the sum of all household members' labor availability.

---

[6] We follow Gao et al. (2011)'s notation for infected ($I_1$) and susceptible individuals ($S_1$). We use similar notation for infected and susceptible snails ($I_2$ and $S_2$), with the subscript 1 for humans and the subscript 2 for snails.

A household generates income by growing food. There are perfectly competitive markets for food, the aggregate household good and urea fertilizer (the tradables set $T = \{f, g, l, u\}$), but there are not markets for vegetation, land or health (the non-tradables set $NT = \{d, v, H\}$). Each household must fully self-provide non-tradable goods. Finally, let $p_i$ denote the market price for good $i$.

Thus, in each period, the household solves the problem:

$$\max_{(\boldsymbol{c},\boldsymbol{q})} U(\boldsymbol{c}, H) \tag{1}$$

subject to the cash budget constraint for tradable goods,

$$p_f c_f + p_g c_g \leq p_f \left( F(L_f, q_d, q_u, \omega q_v) \right) \tag{2}$$

the availability constraint for vegetation use,

$$q_v - \beta_v (L_v)^{\gamma_1} \geq 0 \tag{3}$$

the availability constraint for land use,

$$q_d - e_d \geq 0 \tag{4}$$

the time constraint on the household's labor availability,

$$a_l \geq q_{l_f} + q_{l_v} + c_l \tag{5}$$

and the health production function.

$$H = H(I_1, S_1, c_f) \tag{6}$$

We substitute the availability constraint into the food production function in the cash budget constraint and then substitute the labor constraint into the budget constraint to create the full income constraint:

$$p_f c_f + p_g c_g + w\left(c_l + q_{l_f} + q_{l_v}\right)$$
$$\leq p_f \left( F\left(q_{l_f}, L_f^h, q_d, q_u, \omega q_v(q_{l_v}, L_v^h)\right)\right) - p_u q_u + w a_l \tag{7}$$

Requiring all land to be used in production, assuming an interior solution, substituting (6) into (1) and using Lagrange multiplier $\lambda$ on the household's full income constraint, the first order conditions for the maximization problem are:

$$\frac{\partial U}{\partial c_f} + \frac{\partial U}{\partial H}\frac{\partial H}{\partial c_f} = \lambda p_f \tag{8}$$

$$\frac{\partial U}{\partial c_g} = \lambda p_g \tag{9}$$

$$\frac{\partial U}{\partial c_l} = \lambda w \tag{10}$$

$$p_f \frac{\partial F}{\partial q_{l_f}} = w \tag{11}$$

$$p_f \frac{\partial F}{\partial q_v}\frac{\partial q_v}{\partial q_{l_v}} = w \tag{12}$$

$$p_f \frac{\partial F}{\partial q_u} = p_u \tag{13}$$

Equations (8), (9), and (10) can be rearranged to show that the ratio of the marginal benefit of consuming food (which includes direct increases in utility and indirect utility increases through improved health) to the marginal benefit of consuming the aggregate household good or leisure equals the price ratio. Equations (11) – (13) are input use constraints that require the use of family labor and fertilizer until the value of the marginal product of labor or fertilizer equals its respective cost or opportunity cost in the case of family labor.

Specifically, assume that the household has Cobb-Douglas utility:

$$U(\mathbf{c}, H) = c_f^{\theta_f} c_g^{\theta_g} H^{\theta_h} c_l^{\theta_l} \tag{14}$$

where the $\theta$'s add up to one. We calibrate the parameters $\boldsymbol{\theta}$ by estimating expenditure shares from the Harmonized Survey on Household Living Standards 2018-2019 in Senegal.[12] Expenditure shares can be found in Table S3. We set $\theta_f = 0.5$, $\theta_g = 0.3$, $\theta_h = 0.1$, and $\theta_l = 0.05$.

Health status follows the health production function given by

$$H = \exp\left(\frac{S_1}{I_1 + S_1}\right) c_f^{h_f} \tag{15}$$

where $I_1$ is infected household members, $S_1$ is not infected household members, and $h_f$ is the elasticity of the increase in health from food consumption and we set $h_f = 0.000384$.[17,18]

Production of food takes the CES form:

$$q_f = \left(\alpha_d q_d^\phi + \alpha_l \left(q_{l_f} + L_f^h\right)^\phi + \alpha_u q_u^\phi + \alpha_v (\omega q_v)^\phi\right)^{1/\phi} \tag{16}$$

We estimate factor cost shares from the Harmonized Survey on Household Living Standards 2018-2019 in Senegal to determine the parameters $\alpha_d$, $\alpha_l$, $\alpha_u$, and $\alpha_v$ and calibrate $\phi$ to achieve fertilizer use consistent with observed patterns.[12] Estimated factor cost shares can be found in

Table S4. We set $\alpha_d = 0.4$, $\alpha_l = 0.5$, $\alpha_u = 0.05$, $\alpha_v = 0.05$, and $\phi = 0.3$. We consider labor shares in the model. We scale the production function to labor days based on the average amount of labor allocated to a plot within the survey data as the unit of labor is important for understanding the returns to labor.[19]

We model vegetation harvest as

$$q_v = \beta_v (q_{l_v} + L_v^h)^{\gamma 1} \tag{17}$$

where we set $\beta_v = 14.4942$ and $\gamma 1 = 0.2595$ using estimates of harvested vegetation and labor data from Rohr and colleagues.[11,[7]] We set the price of food, $p_f = 290$ FCFA to the average, location-adjusted price of local rice estimated from Senegalese price reports.[20] We calibrate the price of fertilizer to be consistent with household survey data[12] and to achieve stable aquatic vegetation populations. We set $p_u = 300$ FCFA. We set the price of the aggregate household good to $p_g = 500$ FCFA. In the simulations, we normalize all prices setting the price of food equal to one. A summary of the parameter values used in the household model is presented in Table S5.

*Disease Ecology Model for Schistosomiasis*

The *Ceratophyllum*, the keystone species of aquatic vegetation, population follows a logistic growth function. The population also depends on the amount of vegetation removed by household members or hired workers $q_v$. With a starting density of $N_0$, the population density of aquatic vegetation is

$$\frac{dN}{dt} = r \times N \times \left(1 - \frac{N}{K}\right) + n0 - \frac{q_{v_t}}{365} \tag{18}$$

where $r$ is the net growth rate of *Ceratophyllum*, $K$ is the carrying capacity of the freshwater environment, $n0$ is the recruitment rate of new aquatic vegetation from other parts of the lake or river, and $q_{v_t}$ is the amount of harvested aquatic vegetation, i.e., the household's production of harvested vegetation which is then divided by 365 to model small amounts of daily vegetation harvest by the household. Households harvest vegetation daily as they continuously update their labor allocations consistent with Fafchamps[22] and Dillon (unpublished). The amount of aquatic vegetation to start each period is $N_{t+1} = N_t + \rho \times q_{u_t} \times N_t$ where $\rho \times q_{u_t} \times N_t$ captures the impact of urea fertilizer use, $q_{u_t}$, on vegetation growth as Rohr and colleagues[11,13] reports that agrochemicals like fertilizer contribute to vegetation growth. We estimate the carrying capacity and starting value of *Ceratophyllum* based on the average amount of vegetation found within water access points sampled by Rohr and colleagues,[11] setting $K = 28,906.5\ kg$ and $N_0 = 28,906.5\ kg$. We set $r = 0.05$, $\rho = 0.01$, and $n0 = 0.01$. Table S6 summarizes all parameters in the disease ecology model.

---

[7] Details of the estimation can be found in Appendix A and regression results are in Table 8.

Aquatic vegetation affects the snail population, both susceptible and infected, which we model by

$$\frac{dS_2}{dt} = \Lambda_2 - \frac{\beta_2 M S_2}{M_0 + \epsilon M^2} - (\mu_2 + \chi(K - N))S_2 \tag{19}$$

$$\frac{dI_2}{dt} = \frac{\beta_2 M S_2}{M_0 + \epsilon M^2} - (\mu_2 + \delta_2 + \chi(K - N))I_2 \tag{20}$$

where $\Lambda_2$ is the recruitment rate of susceptible snails, $\beta_2$ is the probability of snail infection from miracidia, $M_0$ is the contact rate between miracidia and snails, $\epsilon$ is the saturation coefficient for miracidial infectivity, $\mu_2$ is the natural death rate of snails, $\delta_2$ is the death rate of snails from infection, and $\chi$ is the death rate of snails from a one kg decrease in vegetation. We set $M_0 = 1 \cdot 0 \times 10^6$, $\epsilon = 0.3$, $\Lambda_2 = 100$, $\beta_2 = 0.615$, $\mu_2 = 0.008$, and $\delta_2 = 0.0004012$,[23] while we estimate $\chi = 0.02842$ from aquatic vegetation removal data.[11, 8]

The miracidia population follows

$$\frac{dM}{dt} = k\lambda_1 I_1 - \mu_3 M \tag{21}$$

where $k$ is the number of eggs released into the environment per human host, $\lambda_1$ is the hatching rate for miracidia, and $\mu_3$ is the miracidial mortality rate. We set $k = 300$, $\lambda_1 = 50$, and $\mu_3 = 2.5$.[23,24] The cercariae population follows

$$\frac{dP}{dt} = \lambda_2 I_2 - \mu_4 P \tag{22}$$

where $\lambda_2$ is the cercarial emergence rate and $\mu_4$ is the cercarial mortality rate. We assume there is no cercarial elimination intervention. We estimate the model with $\lambda_2 = 2.6$ and $\mu_4 = 0.004$.[23]

Finally, the susceptible and infected human populations follow

$$\frac{dS_1}{dt} = -\frac{\beta_1 P S_1}{1 + \alpha_1 P} + \eta I_1 \tag{23}$$

$$\frac{dI_1}{dt} = \frac{\beta_1 P S_1}{1 + \alpha_1 P} - \eta I_1 \tag{24}$$

---

[8] We estimate $\chi$ using a simple calculation comparing the average mass of aquatic vegetation removed at each site to the average drop in snail population after removal.

Where $\beta_1$ is the contact between cercariae and humans, $\alpha_1$ is the saturation coefficient for cercarial infectivity, and $\eta$ is the treatment rate of infected humans. We assume that schistosomiasis infections cause neither birth nor deaths of humans. The unconditional mortality rate of humans due to schistosomiasis is around $\frac{1}{1,000}$.[25] Since we consider villages with average populations around 5,000 with infections around 1,000 – 4,000 at any given time, deaths from schistosomiasis are relatively rare. Thus, we abstract away from the disease's mortality effects and instead focus only on the morbidity impacts through reduced labor productivity. Because we only consider relatively short time periods, we treat the household population as stable and focus on labor availability dynamics within the household. We set $\beta_1 = 1.766 \times 10^{-8}$ and $\alpha_1 = 0.8 \times 10^{-8}$.[23] We set $\eta = 0.0068$ to model some infected individuals receiving treatment through deworming medications (e.g., praziquantel) during sporadic mass deworming events. However, it is expensive to diagnose schistosomiasis and treatment of infections remains relatively infrequent even with mass deworming events that often do not diagnose individuals before they receive deworming medication.

Initial population sizes for all relevant populations in the disease ecology submodel are reported in Table S7.

**SI Appendix Text S3. Additional Information on Estimating Aquatic Vegetation Harvest**

We use experimental field trial data collected from Rohr and colleagues[26] on the amount of vegetation removed and the number of labor days devoted to harvesting vegetation to estimate the parameters in the production function of harvested vegetation (Equation 17). We estimate the harvested vegetation production as

$$\ln(Kg\ of\ harvested\ vegetation)_i = \alpha + \beta \ln(person\ days)_i + \varepsilon_i \qquad (25)$$

The coefficient estimate $\beta$ is our direct estimate of $\gamma 1$ in Equation 19 and we calculate $\beta_v$ from the estimate of the constant $\alpha$ using $\beta_v = \exp(\alpha)$. Results from the estimation are reported in Table S8.

**SI Appendix Text S4. Additional Details on Disease Ecology Submodel Parameterization**

We base the disease ecology submodel on Gao and colleagues.[23] We use experimental estimates of parameters in the local population in Senegal from Nguyen and colleagues[24] as a guide to adjust model parameters to match human infection levels observed within the region. Table S9 reports and describes the starting parameters we used to simulate the model. We excluded human births and deaths from this simulation.[9]

The continuous time equations are:

Susceptible snails:

$$\frac{dS_2}{dt} = \Lambda_2 - \frac{\beta_2 M S_2}{M_0 + \epsilon M^2} - \mu_2 S_2 \qquad (26)$$

Infected snails:

$$\frac{dI_2}{dt} = \frac{\beta_2 M S_2}{M_0 + \epsilon M^2} - (\mu_2 + \delta_2) I_2 \qquad (27)$$

Cercariae:

$$\frac{dP}{dt} = \lambda_2 I_2 - \mu_4 P \qquad (28)$$

Susceptible Humans:

$$\frac{dS_1}{dt} = -\frac{\beta_1 P S_1}{1 + \alpha_1 P} + \eta I_1 \qquad (29)$$

Infected Humans:

$$\frac{dI_1}{dt} = \frac{\beta_1 P S_1}{1 + \alpha_1 P} - \eta I_1 \qquad (30)$$

---

[9] Over a relatively short time horizon, 20 years or less, assuming away human population growth or decline for an individual family is reasonable as it represents roughly one generation.

Miracidia:

$$\frac{dM}{dt} = k\lambda_1 I_1 - \mu_3 M \tag{31}$$

*Modifications*

We calibrated the human population to match the household-level analysis in the Senegalese context. The household size is set at 10 where 7.5 humans start as susceptible and 2.5 are infected, matching the 25% baseline prevalence of *S. mansoni* in the region reported by Rohr and colleagues.[11] Modifications to the original model parameters reported in Gao and colleagues[23] are required because we significantly reduce the size of the human population and eliminate human births and deaths to integrate the disease ecology model of schistosomiasis with an economic model of agricultural households.

We start with the parameters in Gao and colleagues[23] and then calibrate the model from these parameter starting points with the goal of finding a steady state at or very close to 25% infection with 10 humans in the model (so 7.5 susceptible humans and 2.5 infected humans). We calibrate the parameters to achieve population stability in the snails and then increase infection until the human infection stabilized near 25%.

Finally, we added vegetation into the model. We use a general logistic growth function for vegetation, where $r$ is the growth rate, $K$ is the carrying capacity, and $n0$ is the natural recolonization rate. The carrying capacity was estimated from vegetation data.[11] We chose the growth rate and the recolonization rate to match rapid regrowth consistent with rates observed at study sites in Rohr and colleagues.[11] The logistic growth function is reported below:

$$\frac{dN}{dt} = r \times N \times \left(1 - \frac{N}{K}\right) + n0 \tag{32}$$

To connect vegetation to the existing system, a parameter $\chi$ is added to the snails' population equations. For every kilogram of vegetation below the carrying capacity, the snail population is reduced by $\chi$ percent. We start the vegetation population at the carrying capacity and do not include vegetation removal and thus vegetation has no effect on the other populations in these model runs. Table S10 reports all starting values and adjusted parameters.

*Simulations*

Results from the simulations for each of key populations can be found in Figure S11. We present five-year models of simulations without vegetation to confirm we have found a steady state within the disease ecology submodel. Since the vegetation population is started at the steady state level, it does not affect how the rest of the model operates and thus is not needed in these extra simulations to confirm the snails, humans, miracidia, and cercariae populations approach a steady state.

**Supplemental Tables**

**Table S1. Land endowments for household simulations.** Land holdings based on the 25th, 50th, and 75th percentiles in the Saint Louis and Louga regions from the Harmonized Survey on Household Living Standards in Senegal collected in 2018 and 2019.

| Type | Land Endowment (hectares) |
| --- | --- |
| 25th percentile | 0.5 |
| 50th percentile | 2 |
| 75th percentile | 5.5 |

**Table S2. Summary statistics of agricultural households in the Saint Louis and Louga regions.** Summary statistics for households in the Saint Louis and Louga regions of the 2018-2019 Harmonized Survey on Household Living Standards in Senegal. Household size is calculated by summing the number of household members included in the member module of the survey. Household farm labor and outside labor includes labor of all household members across the following tasks: preparing the plot, weeding, and harvesting. Female indicates that the household head is female. Read French and Write French indicate that the household head can read or write in French, respectively. Formal school indicates that the household head attended formal schooling. Hire outside labor indicates that the household hired at least one person day of labor from an individual outside the family. Rice, Millet, Cowpea, and Peanut indicates that the household in engaged in rice, millet, cowpea, or peanut cultivation, respectively. Irrigation and Fertilizer indicate that at least one household plot is irrigated or uses fertilizer, respectively.

|  | N | Mean | St. Dev. | Min | Max |
|---|---|---|---|---|---|
| *Household Head* | | | | | |
| Female (1 = yes) | 984 | 0.287 | 0.452 | 0 | 1 |
| Age (years) | 984 | 52.725 | 14.269 | 20 | 95 |
| Married (1 = yes) | 984 | 0.854 | 0.354 | 0 | 1 |
| Read French (1 = yes) | 983 | 0.312 | 0.464 | 0 | 1 |
| Write French (1 = yes) | 983 | 0.306 | 0.461 | 0 | 1 |
| Formal School (1 = yes) | 983 | 0.304 | 0.460 | 0 | 1 |
| *Household* | | | | | |
| Household Size (persons) | 984 | 10.643 | 6.675 | 1 | 58 |
| Household Farm Labor (person days) | 384 | 388.672 | 462.713 | 0 | 2909 |
| Hire Outside Labor (1 = yes) | 384 | 0.430 | 0.496 | 0 | 1 |
| Outside Labor (person days) | 394 | 23.388 | 55.696 | 0 | 348 |
| Rice (1 = yes) | 384 | 0.273 | 0.446 | 0 | 1 |
| Millet (1 = yes) | 384 | 0.242 | 0.429 | 0 | 1 |
| Cowpea (1 = yes) | 384 | 0.474 | 0.500 | 0 | 1 |
| Peanut (1 = yes) | 384 | 0.466 | 0.500 | 0 | 1 |
| Irrigation (1 = yes) | 384 | 0.378 | 0.485 | 0 | 1 |
| Fertilizer (1 = yes) | 378 | 0.457 | 0.499 | 0 | 1 |

**Table S3. Estimated expenditure shares.** Estimated expenditure shares from the Harmonized Survey on Household Living Standards in Senegal collected in 2018 and 2019. We classified goods according to three categories: food, health, and household goods where household goods captured goods that did not clearly fit into food or health. We then aggregated annual expenditure for each of the goods in these categories. Some expenditures recorded in the survey were excluded, therefore the totals may not add up to 1.[10] Fewer households report cash health expenditures, so we take these expenditure share estimates as a lower bound when calibrating the household's utility function focusing on the expenditure share estimates for food and household goods.

|  | N | Mean | St. Dev. | Min | Max |
|---|---|---|---|---|---|
| Food Expenditure Share | 7156 | 0.539 | 0.131 | 0.027 | 0.941 |
| Household Good Expenditure Share | 7156 | 0.313 | 0.126 | 0.007 | 0.971 |
| Health Expenditure Share | 6035 | 0.036 | 0.052 | 0 | 0.798 |

---

[10] We exclude alcohol and tobacco purchases. Since we abstract away from the land market, we exclude any payments for land or housing.

**Table S4. Estimated factor cost shares.** Estimated factor cost shares from the Harmonized Survey on Household Living Standards in Senegal collected in 2018 and 2019. We measure land in hectares and then valued land using the rental price of 120,000 FCFA per hectare as reported in the Saint Louis region. We then calculated a household's total labor days on each plot by the following tasks: prepping the land, weeding, and harvesting. We include both family and hired labor and then calculate total labor by adding up all the labor days on each of the family's plots including all three tasks. We then use the median adult male harvesting wage in each region as the value of each day of labor to calculate the total cost of labor. Inorganic fertilizer includes urea, NPK, and phosphates and is measured in kgs. We use the median regional price for each type of inorganic fertilizer when calculating the factor cost. Compost is also measured in kgs. As with inorganic fertilizer, we use the median regional price for animal compost to calculate the factor cost. All carts and sacs are assumed to be 50 kg of fertilizer or animal compost.

|  | N | Mean | St. Dev. | Min | Max |
|---|---|---|---|---|---|
| Land Factor Cost Share | 2892 | 0.416 | 0.257 | 0 | 1 |
| Labor Factor Cost Share | 2892 | 0.529 | 0.265 | 0 | 0.999 |
| Inorganic Fertilizer Factor Cost Share | 2892 | 0.037 | 0.091 | 0 | 0.990 |
| Compost Factor Cost Share | 1277 | 0.040 | 0.084 | 0 | 1 |

**Table S5. Parameters for the household model.** The $\theta$ parameters for the utility function are based on household expenditure share estimates from the Saint Louis and Louga regions in the Harmonized Survey on Household Living Standards reported in Table S3. We round the expenditure share estimates for food and household goods and then scale the parameters on health status and leisure so that the sum of all $\theta$'s adds up to one. The parameter $h_f$ is taken from Pitt and colleague's[18] estimate of the relationship between caloric intake and health. We scale the estimate to fit our measure of calories in one kg of rice which is the unit of food in the model. The $\alpha$ parameters for the food production function are based off of factor cost share estimates from the Saint Louis and Louga regions in the Harmonized Survey on Household Living Standards reported in Table S4. We round the factor cost share estimates so that the $\alpha$'s add up to one. The substitution parameter $\phi$ is calibrated to achieve fertilizer use levels consistent with the Senegalese context. The mass loss during compost, modeled by the parameter $\omega$, is based on the range of estimates in Şevik and colleagues[21] and calibrated to achieve fertilizer use consistent with the Senegalese context. The parameters $\beta_v$ and $\gamma 1$ are estimated from data on vegetation removal done in Rohr and colleagues[11] and reported in SI Appendix Text S3. The price of food comes from Senegalese price reports released by ANSD and the price of fertilizer is consistent with the Harmonized Survey on Household Living Standards and calibrated to achieve stability in the simulations. The price of the household good is calibrated to capture the value of many possible consumption goods the household purchases which are more expensive than food.

| Parameter | Description | Value |
|---|---|---|
| $\theta_f$ | Utility function coefficient on food | 0.55 |
| $\theta_g$ | Utility function coefficient on household goods | 0.3 |
| $\theta_h$ | Utility function coefficient on health status | 0.1 |
| $\theta_l$ | Utility function coefficient on leisure | 0.05 |
| $h_f$ | Coefficient on food consumption in health status function | 0.000384 |
| $\alpha_d$ | Coefficient on land in food production | 0.4 |
| $\alpha_l$ | Coefficient on labor in food production | 0.5 |
| $\alpha_u$ | Coefficient on fertilizer in food production | 0.05 |
| $\alpha_v$ | Coefficient on vegetation in food production | 0.05 |
| $\omega$ | Vegetation retained in composting | 0.6 |
| $\phi$ | Substitution parameter | 0.3 |
| $\beta_v$ | Coefficient for harvesting vegetation | 14.4942 |
| $\gamma 1$ | Exponent on labor in harvesting vegetation | 0.2595 |
| $p_f$ | Price of food | 290 FCFA |
| $p_h$ | Price of household good | 500 FCFA |
| $p_u$ | Price of fertilizer | 300 FCFA |

**Table S6. Parameters for the disease ecology model.** The parameters $\beta_2$, $\mu_4$, $\lambda_2$, $M_0$, $\epsilon$, and $k$ are from Gao and colleagues.[23] The parameters $\Lambda_2$, $\mu_2$, and $\delta_2$ are calibrated to achieve a stable snail population throughout the simulations. The parameters $\lambda_1$ and $\mu_3$ are calibrated to achieve a stable miracidia population throughout the simulations. The parameters $\beta_1$ and $\alpha_1$ are calibrated to achieve stable infection rates in humans consistent with the 25% infection rate from data collected by Rohr and colleagues.[11] The parameters $K$ and $\chi$ are estimated from data collected by Rohr and colleagues.[11] The parameters $r$, $\rho$, and $n0$ are calibrated to fit the high growth rate of vegetation observed in the Senegalese context and to adequately capture the effect of fertilizer runoff on vegetation growth. $\eta$ is calibrated to deworming every four years.

| Parameter | Description | Value |
| --- | --- | --- |
| $r$ | Vegetation growth rate | 0.05 |
| $K$ | Vegetation carrying capacity | 28,906.5 kg |
| $\rho$ | Effect of fertilizer on vegetation growth | 0.01 |
| $n0$ | Vegetation recolonization rate | 0.01 |
| $\Lambda_2$ | Snail recruitment rate | 100 |
| $\beta_1$ | Contact between cercariae and humans | $1.766 \times 10^{-8}$ |
| $\beta_2$ | Probability of snail infection from miracidia | 0.615 |
| $\mu_2$ | Snail natural mortality rate | 0.008 |
| $\mu_3$ | Miracidial mortality rate | 2.5 |
| $\mu_4$ | Cercarial mortality rate | 0.004 |
| $\delta_2$ | Snail death rate from infection | 0.0004012 |
| $\lambda_1$ | Hatching rate of miracidia | 50 |
| $\lambda_2$ | Cercarial emergence rate | 2.6 |
| $\alpha_1$ | Saturation coefficient for cercarial infectivity | $0.8 \times 10^{-8}$ |
| $M_0$ | Contact rate between miracidia and snails | $1.00 \times 10^6$ |
| $\epsilon$ | Saturation coefficient for miracidial infectivity | 0.30 |
| $\chi$ | Snail death rate from vegetation removal | 0.02842 |
| $k$ | Eggs released per infected human | 300 |
| $\eta$ | Treatment rate of infected humans | 0.00068 |

**Table S7. Starting values of the disease ecology populations.** Average household size begins at the nearest whole number with easy division into 4 of 10 based on the average household size in the Saint Louis and Louga region from the Harmonized Survey on Household Living Standards in Senegal, 2018-2019. Infected and susceptible humans were then calculated based on the average infection prevalence of *S. mansoni* in the infection data from Rohr and colleagues.[11] All other parameters were calibrated to be consistent with the human infection data.

| Parameter | Description | Value |
|---|---|---|
| $N_0$ | Starting amount of vegetation | 28,906.5 kg |
| $S_1$ | Susceptible humans | 7.5 |
| $I_1$ | Infected humans | 2.5 |
| $S_2$ | Susceptible snails | 200 |
| $I_2$ | Infected snails | 12,300 |
| $M$ | Miracidia | 15,000 |
| $P$ | Cercariae | 130,000 |

**Table S8. Vegetation production function estimates.** Estimates of the vegetation production function in equation 19. Huber-White robust standard errors in parentheses. * $p < 0.1$, ** $p < 0.05$, *** $p < 0.01$

|  | Log(kg of vegetation) |
|---|---|
| Log(person days) | 0.260*** |
|  | (0·0581) |
| Constant | 2.674*** |
|  | (0.141) |
| N | 92 |
| Adj. $R^2$ | 0.208 |

**Table S9. Parameters for the disease ecology model in Gao and colleagues.**[23] Parameter values are taken directly from Gao and colleagues[23] and reported here.

| Parameter | Description | Value |
|---|---|---|
| $\Lambda_2$ | Snail recruitment rate | 200 d$^{-1}$ |
| $\beta_1$ | Contact between cercariae and humans | $0.406 \times 10^{-8}$ |
| $\beta_2$ | Probability of snail infection from miracidia | 0.615 |
| $\mu_2$ | Snail natural mortality rate | 0.000569 |
| $\mu_3$ | Miracidial mortality rate | 0.9 |
| $\mu_4$ | Cercarial mortality rate | 0.004 |
| $\delta_2$ | Snail death rate from infection | 0.0004012 |
| $\lambda_1$ | Hatching rate of miracidia | 0.00232 |
| $\lambda_2$ | Cercarial emergence rate | 2.6 |
| $\alpha_1$ | Saturation coefficient for cercarial infectivity | $0.3 \times 10^{-8}$ |
| $M_0$ | Contact rate between miracidia and snails | $1.00 \times 10^6$ |
| $\epsilon$ | Saturation coefficient for miracidial infectivity | 0.30 |
| $k$ | Eggs released per infected human | 300 |
| $\eta$ | Treatment rate of infected humans | 0.00068 |

**Table S10. Adjusted parameters for the disease ecology model.** A value of "Yes" in the modification column reports that the parameter used in the disease ecology model differs from the model reported in Gao and colleagues.[23] Vegetation does not exist in the Gao and colleagues' model so all vegetation parameters are modifications.

| Parameter | Description | Value | Modification |
|---|---|---|---|
| $r$ | Vegetation growth rate | 0.05 | Yes |
| $K$ | Vegetation carrying capacity | 28,906.5 kg | Yes |
| $\rho$ | Effect of fertilizer on vegetation growth | 0.01 | Yes |
| $n0$ | Vegetation recolonization rate | 0.01 | Yes |
| $\Lambda_2$ | Snail recruitment rate | 100 | Yes |
| $\beta_1$ | Contact between cercariae and humans | $1.766 \times 10^{-8}$ | Yes |
| $\beta_2$ | Probability of snail infection from miracidia | 0.615 | No |
| $\mu_2$ | Snail natural mortality rate | 0.008 | Yes |
| $\mu_3$ | Miracidial mortality rate | 2.5 | Yes |
| $\mu_4$ | Cercarial mortality rate | 0.004 | No |
| $\delta_2$ | Snail death rate from infection | 0.0004012 | Yes |
| $\lambda_1$ | Hatching rate of miracidia | 50 | Yes |
| $\lambda_2$ | Cercarial emergence rate | 2.6 | No |
| $\alpha_1$ | Saturation coefficient for cercarial infectivity | $0.8 \times 10^{-8}$ | Yes |
| $M_0$ | Contact rate between miracidia and snails | $1.00 \times 10^6$ | No |
| $\epsilon$ | Saturation coefficient for miracidial infectivity | 0.30 | No |
| $\chi$ | Snail death rate from vegetation removal | 0.02842 | Yes |
| $k$ | Eggs released per infected human | 300 | No |
| $\eta$ | Treatment rate of infected humans | 0.00068 | No |

**Supplemental Figures**

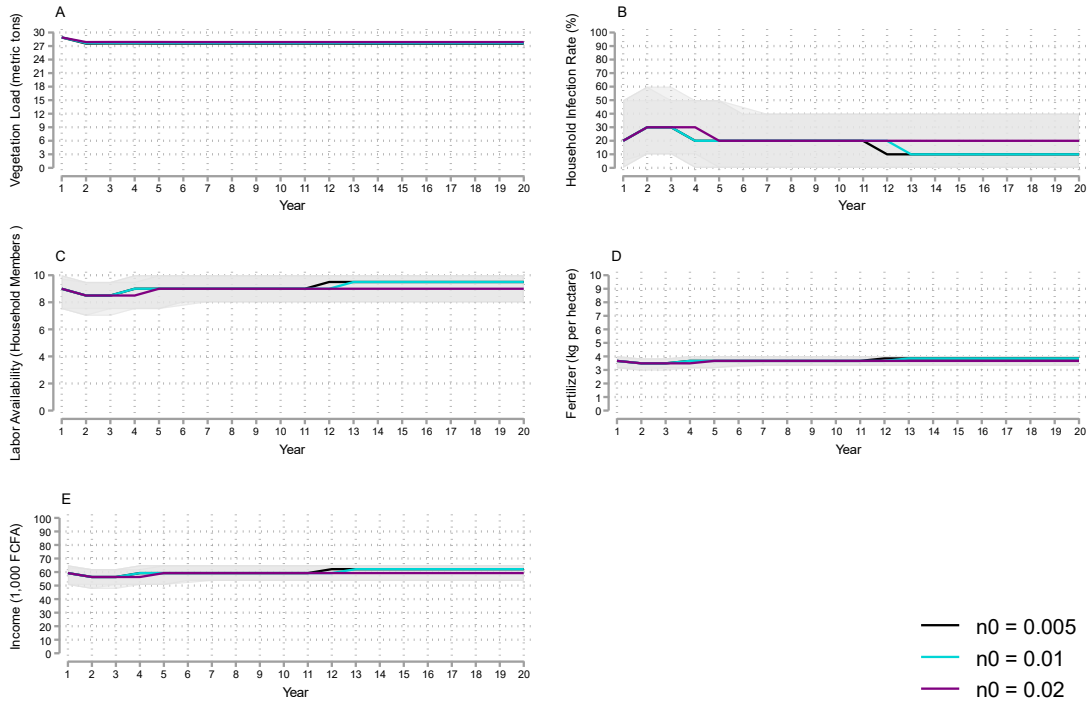

**Figure S1. Median vegetation load, infection rate, labor availability, fertilizer use, and income for the recolonization rate sensitivity analysis.** Panel A (top left) plots the median aquatic vegetation stock (population) in metric tons across 1,000 20-year simulations for three different levels of the vegetation recolonization rate ($n0$) and households with two hectares of land. Aquatic vegetation load represents the size of the snail habitat within the village water access point used by the household. Shaded areas represent the 5-95 percent centered confidence band. Based on scale and precision, not all shaded areas are visible. Panel B shows the median household infection rate (the number of infected individuals divided by total number of household members). Panel C reports median labor availability from the 10-person household size maximum. Panel D displays median fertilizer use in kgs per hectare, and Panel E reports the median income in FCFA1,000. Medians and percentiles are within each land endowment each time period across the 1,000 simulations.

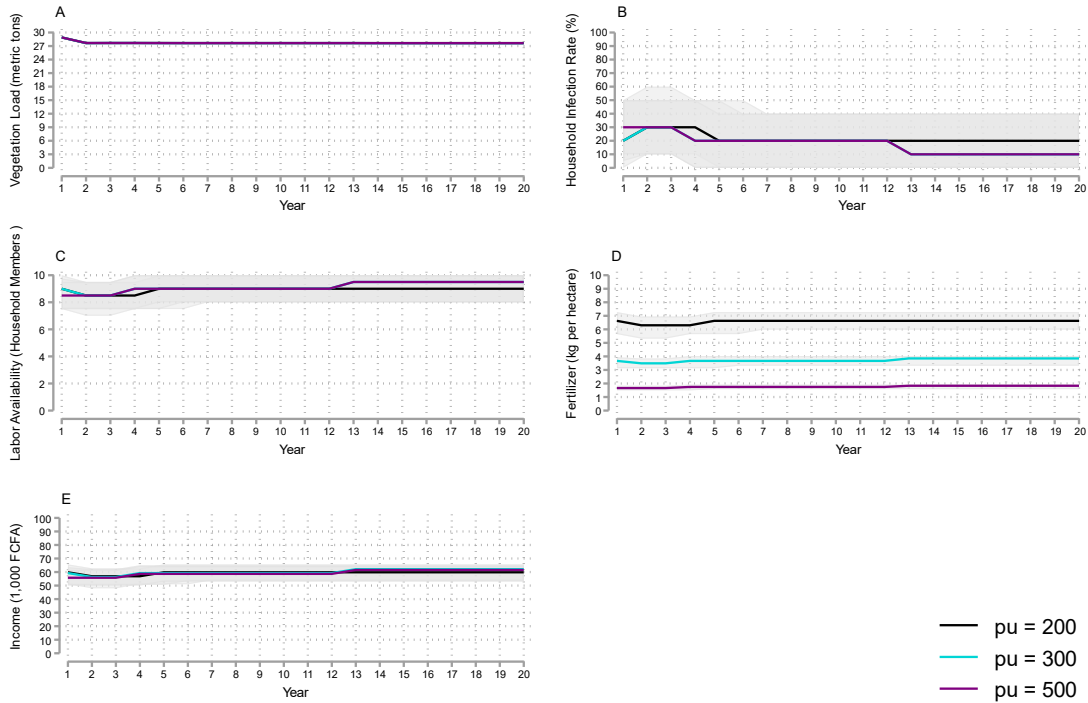

**Figure S2. Median vegetation load, infection rate, labor availability, fertilizer use, and income for the fertilizer price sensitivity analysis.** Panel A (top left) plots the median aquatic vegetation stock (population) in metric tons across 1,000 20-year simulations for three different levels of fertilizer prices ($p_u$) and households with two hectares of land. Aquatic vegetation load represents the size of the snail habitat within the village water access point used by the household. Shaded areas represent the 5-95 percent centered confidence band. Based on scale and precision, not all shaded areas are visible. Panel B shows the median household infection rate (the number of infected individuals divided by total number of household members). Panel C reports median labor availability from the 10-person household size maximum. Panel D displays median fertilizer use in kgs per hectare, and Panel E reports the median income in FCFA1,000. Medians and percentiles are within each land endowment each time period across the 1,000 simulations.

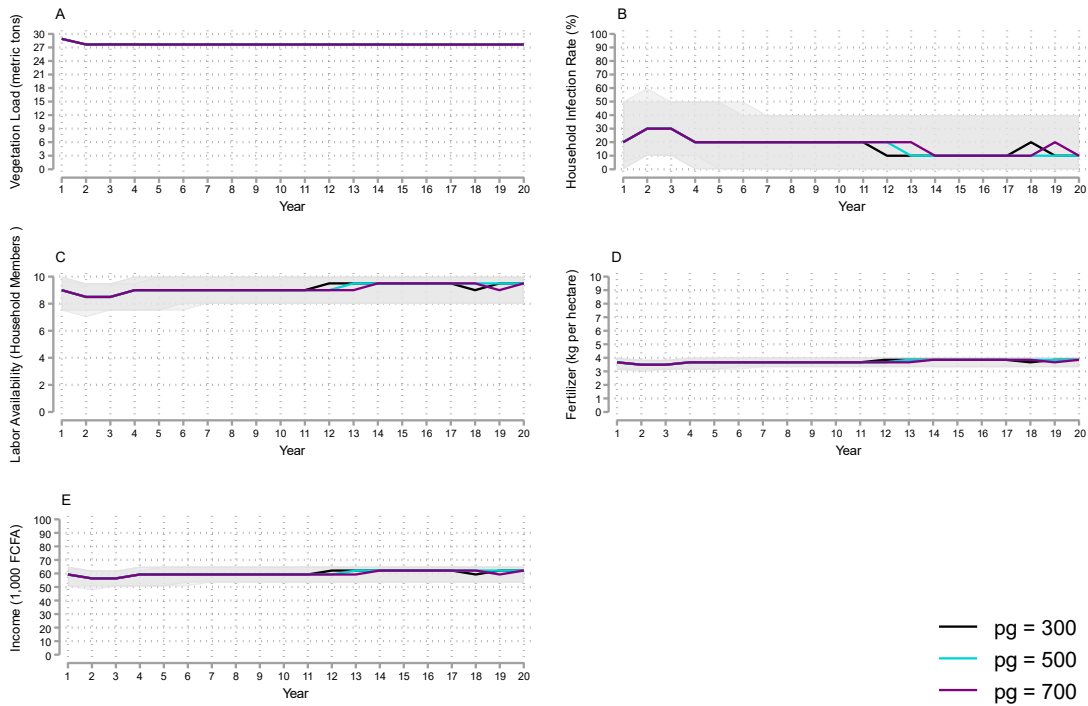

**Figure S3. Median vegetation load, infection rate, labor availability, fertilizer use, and income for the household good price sensitivity analysis.** Panel A (top left) plots the median aquatic vegetation stock (population) in metric tons across 1,000 20-year simulations for three different levels of household good prices ($p_h$) and households with two hectares of land. Aquatic vegetation load represents the size of the snail habitat within the village water access point used by the household. Shaded areas represent the 5-95 percent centered confidence band. Based on scale and precision, not all shaded areas are visible. Panel B shows the median household infection rate (the number of infected individuals divided by total number of household members). Panel C reports median labor availability from the 10-person household size maximum. Panel D displays median fertilizer use in kgs per hectare, and Panel E reports the median income in FCFA1,000. Medians and percentiles are within each land endowment each time period across the 1,000 simulations.

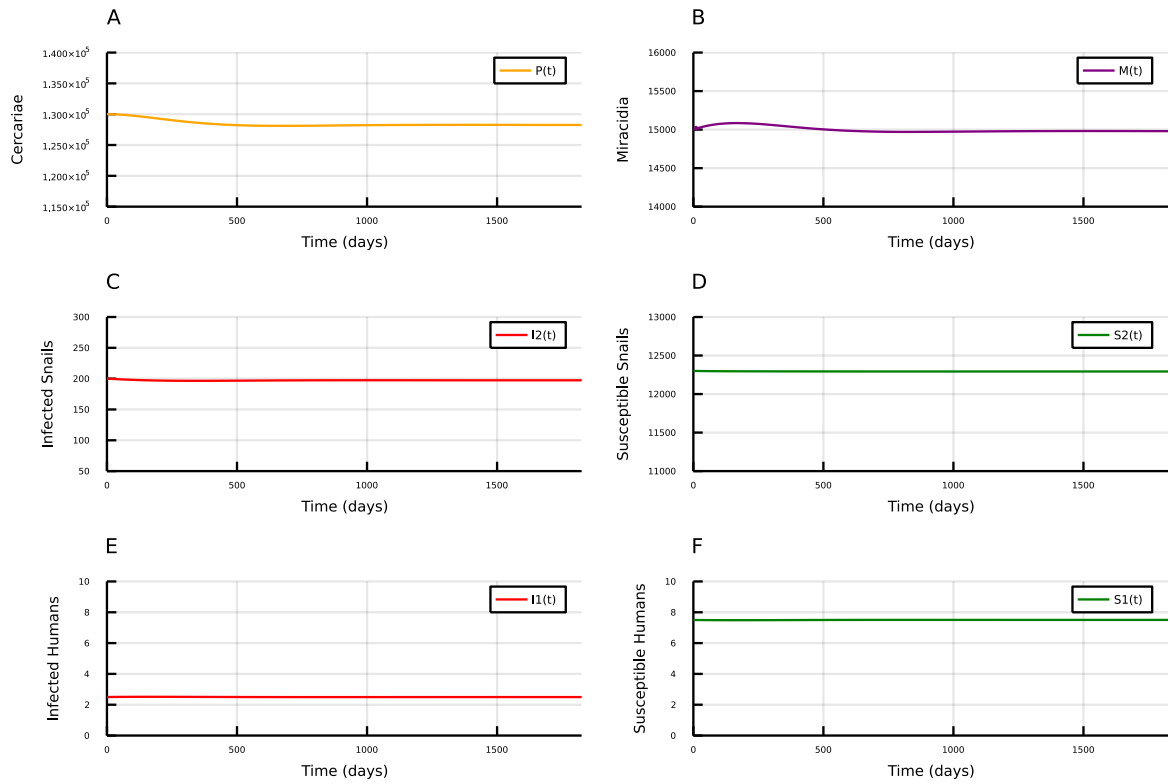

**Figure S4. Five-year continuous time simulation results.** Simulation results are for the modified disease ecology model. Vegetation is omitted as it is set to the carrying capacity and has no effect on the system in this stable state.